\documentclass[prb,floatfix,aps,showpacs,preprintnumbers,amsmath,amssymb]{revtex4}
\usepackage{natbib}
\usepackage{amsmath,amssymb}
\usepackage{graphicx}
\usepackage{subfigure}
\usepackage{times}
\usepackage{textcomp}
\begin{document}

\title{Hyperfine Interactions in MnAs studied by Perturbed Angular
Correlations of $\gamma$-Rays using the probe $^{77}$Br$\rightarrow^{77}$Se and first
principles calculations for MnAs and other Mn pnictides}
\author{J. N. Gon\c{c}alves}
\email{joaonsg@ua.pt}
\author{V. S. Amaral}
\affiliation{Departmento de F\'isica and CICECO, Universidade de Aveiro, 3810-193 Aveiro, Portugal}
\author{J. G. Correia}
\affiliation{Instituto Tecnol\'ogico e Nuclear, UFA, 2686-953 Sacav\'em, Portugal}
\author{A. M. L. Lopes}
\affiliation{Centro de F\'isica Nuclear da Universidade de Lisboa, 1649-003 Lisboa, Portugal}

\date{\today}

\begin{abstract}
The MnAs compound shows a first-order transition at T$_C\approx42$ C, 
and a second-order transition at T$_t\approx120$ C. The first-order 
transition, with structural (hexagonal-orthorhombic), magnetic (FM-PM) 
and electrical conductivity changes, is associated to magnetocaloric, 
magnetoelastic, and magnetoresistance effects. 
We report a study in a large temperature range from $-196$ up to $140$ C, using 
the  $\gamma-\gamma$ perturbed angular correlations method with the 
radioactive probe $^{77}$Br$\rightarrow^{77}$Se, produced at the 
ISOLDE-CERN facility.
The electric field gradients and magnetic hyperfine fields are determined 
across the first- and second-order phase transitions encompassing the 
pure and mixed phase regimes in cooling and heating cycles.
The temperature irreversibility of the 1st order phase transition is 
seen locally, at the nanoscopic scale sensitivity of the hyperfine 
field, by its hysteresis, detailing and complementing information 
obtained with macroscopic measurements (magnetization 
and X-ray powder diffraction). 
To interpret the results, hyperfine parameters were obtained with 
first-principles spin-polarized density functional calculations using 
the generalized gradient approximation with the full potential (L)APW+lo 
method (\textsc{Wien2k} code) by considering the Se probe at both Mn and 
As sites. A clear assignment of the probe location at the As site is made 
and complemented with the calculated densities of states and local magnetic 
moments. We model electronic and magnetic properties of the chemically 
similar MnSb and MnBi compounds, complementing previous calculations.
\end{abstract}

\maketitle
%%%%%%%%%%%%%%%%%%%%%%%%%%%%%%%%%%%%%%%%%%%%%%%%%%%%%%%%%%%%%%%%%%%%%
%%%%%%%%%%%%%%%%%%%%%%%%%%%%%%%%%%%%%%%%%%%%%%%%%%%%%%%%%%%%%%%%%%%%%
\section{Introduction}
%%%%%%%%%%%%%%%%%%%%%%%%%%%%%%%%%%%%%%%%%%%%%%%%%%%%%%%%%%%%%%%%%%%%%
%%%%%%%%%%%%%%%%%%%%%%%%%%%%%%%%%%%%%%%%%%%%%%%%%%%%%%%%%%%%%%%%%%%%%

The magnetic compound MnAs has been intensively studied, since 
it exhibits a magnetocaloric effect~\cite{MnAs:Wad2001}, under 
hydrostatic pressure~\cite{PhysRevLett.93.237202}, as well as when
doped with metals~\cite{MnAs:Wad2001,NatureMnAs}, making 
it an interesting
material for magnetic refrigeration applications.
Moreover, it can be grown
as epitaxial films on Si and GaAs substrates~\cite{daweritz}, where
applications such as a source for spin injection make it of
promising use for spintronics~\cite{PhysRevB.66.081304}.

In parallel, it is a material with theoretical challenges. 
In this front some
first-principles studies are directed to this compound, e.\ g. see 
refs.~\cite{MnAs:San2006,MnAs:Run2006,MnAs:Nir2004, MnAs:Rav99,PhysRevLett.104.147205}. 
The orthorhombic phase is usually considered paramagnetic, 
however it does not follow a Curie-Weiss law and it has also been considered 
to be antiferromagnetic~\cite{MnAs:Nir2004}. Also of interest 
is the existing magnetoresistance
effect which is attempted to be related to the CMR found in
the perovskite manganites~\cite{MnAs:Mir2003}, and a 
remarkable spin-phonon coupling found crucial to the 
magnetostructural transition~\cite{PhysRevLett.104.147205}.
Its particular coupling of magnetism and structure has been 
the origin of  macroscopic models~\cite{beanrodbell} for magneto-volume effects. 

At low temperatures, MnAs is
fer\-roma\-gne\-tic and it has a
NiAs-type structure. This stru\-cture%or $B8_{1}$ phase
, with space group $P6_3mmc$ (194),
has Mn and As atoms at co\-ordinates ($0$,$0$,$0$) and ($1/3$, $2/3$, $1/4$), 
res\-pectively,
with two formula units per unit cell. On heating, at about $40$ C, it 
un\-der\-goes a first-order phase transition,
with a discontinuous distortion to the orthorhombic MnP-type 
structure,
with a parallel dis\-con\-tinuous change of volume, loss of 
ferroma\-gnetism, and a metal-insulator transition. The orthorhombic 
distortion continuously disappears when heating
until about $125$ C where it undergoes a second-order phase transition to the
NiAs-type structure with a paramagnetic state, now following a Curie-Weiss law. 

Thermal hysteresis is measured in this transition: on heating, 
the hexagonal$\rightarrow$orthorhombic phase transformation 
occurs at temperatures T$_{C,i}\approx 40.5-42.5$ C, while on 
cooling this transformation occurs at T$_{C,d}\approx 33.9-37.9$ 
C~\cite{MnAs:Goo67,Nascimento2006,MnAs:Wil64,JJAP.42.L918,Ishikawa2004408} 
(variations in different studies are probably resulting from small 
differences in the stoichiometry of samples). 
Phase coexistence in a temperature interval of 
approximately $2$ C is reported
by neutron and X-ray diffraction 
measurements~\cite{MnAs:Mir2003,Gama2009}.
The orthorhombic $B31$ structure, with the space group $Pnma$(62),
has coordinates for Mn and As atoms of ($0.995$, $1/4$, $0.223$) 
and ($0.275$, $1/4$, $0.918$), respectively~\cite{MnAs:Wil64}.

The lattice constants of the hexagonal phase at room 
temperature are $a=3.722$, $c=5.702$ \AA{},
and changes to $a=5.72$, $b=3.676$ and $c=6.379$ \AA{} in 
the orthorhombic phase (with four f. u. per unit cell) 
at the first-order transition correspond to a volume loss of 2\%.

We report a study in this compound using $\gamma-\gamma$
time differential Perturbed Angular Correlation (PAC) 
spectroscopy (see e.\ g.~\cite{Schatz} for details),
to our knowledge the first use of this nuclear technique in the compound. 
Since PAC measures the combined hyperfine interactions - 
magnetic hyperfine field (MHF), and electric field gradient (EFG), the 
sensitivity of its atomic-scale measurements allows 
the study of the atomic environments as a function of temperature.

Other hyperfine interactions techniques have been used for the 
study of MnAs and related compounds in previous studies. 
Using M\"{o}ssbauer spectroscopy with the $^{57}$Fe probe at 0.25 at\% concentration, 
Kirchschlager et al.~\cite{Kirchschlager1981} detect a quadrupole 
splitting, which they interpret on the basis of motion of the 
probe atoms, but they do not measure magnetic hyperfine field. 
Also using $^{57}$Fe impurities as probes, in the 
related MnAs$_{1-x}$Fe$_x$ compound~\cite{Abdelgadir1988}, 
with $x=0.01$, $0.03$, and $0.15$, Abdelgadir et al. reported 
measurements involving the first-order transition at $T_{C,d}=2$ 
C (for $x=0.01$), where they also detect an unusual 
dependence of the magnetic hyperfine field. 
NMR spectroscopy has also been performed at 4K~\cite{Amako}, and in the range 
from $-190$ up to $38$ C, with double signals from both Mn 
and As atoms, where a resonance anomaly was observed  
at $\approx -5$0 C as due from atoms at the domain walls~\cite{PINJARE1982}.

Our work studies a temperature range from $13$ to $140$ C and 
liquid nitrogen temperature ($-196$ C). Measurements 
are made in the first-order phase-transition region
and above, passing the second-order phase transition (section~\ref{firstex}).
The temperature range near the first-order transition 
is studied in more detail (section~\ref{secondex}). X-Raw powder 
diffraction and magnetization measurements are also performed 
and its results are compared with PAC results.

The experimental results  are complemented with density 
functional theory calculations of the hyperfine parameters, 
using the  full potential mixed (linear) augmented plane 
wave plus local orbitals (L)APW+lo  method.
In order to improve and complement other first-principles studies,
we also show calculations of other properties,
and for the chemically similar manganese pnictides MnSb and MnBi.

%%%%%%%%%%%%%%%%%%%%%%%%%%%%%%%%%%%%%%%%%%%%%%%%%%%%%%%%%%%%%%%%%%%
%%%%%%%%%%%%%%%%%%%%%%%%%%%%%%%%%%%%%%%%%%%%%%%%%%%%%%%%%%%%%%%%%%%
\section{Experiments}
%%%%%%%%%%%%%%%%%%%%%%%%%%%%%%%%%%%%%%%%%%%%%%%%%%%%%%%%%%%%%%%%%%%
%%%%%%%%%%%%%%%%%%%%%%%%%%%%%%%%%%%%%%%%%%%%%%%%%%%%%%%%%%%%%%%%%%%
\subsection{Hyperfine Parameters}
The two quantities of interest to the physics of MnAs that
can be obtained from the PAC measurements are the electric 
field gradient (EFG) and the magnetic hyperfine field (B$_{hf}$).
The EFG is measured from the hyperfine interaction between a
charge distribution with non-spherical sy\-mmetry and the 
nuclear quadrupole moment Q.
The measurement of the quadrupole interaction gives the EFG, 
depending on the accu\-rate knowledge of the probe's quadrupolar moment.
The EFG is defined as the symmetric traceless tensor with components
taken from the second spatial derivatives of the
Coulomb potential at the nuclear position.
In the principal axis frame of reference,
the components of interest are V$_{zz}$ and the axial symmetry parameter $\eta$,
with $|V_{zz}| > |V_{yy}| \ge |V_{xx}|$ and $\eta = (V_{xx} - V_{yy})/V_{zz}$.
The observable frequency with PAC depends on the quadrupole
nuclear moment and electric field gradient in the following way:
\begin{equation}
\label{nuQ}
\omega_\phi=\frac{2\pi}{4I(2I-1)}\nu_Qk\textnormal{, with } \nu_Q=\frac{eQV_{zz}}{h}\textnormal{, for } \eta=0,
\end{equation}
where $I$ is the nuclear spin and $k=6$ for half-integer spin. $\nu_Q$ is 
called the ``reduced frequency'' of the interaction and is 
independent of $\eta$ and $I$.

The magnetic hyperfine field, arising from the
dipole-dipole interaction between the nuclear magnetic moment
and the magnetic moment of the extranuclear electrons, can be expressed by
\begin{equation}
\label{bhf}
B_{hf}=\frac{\omega_L\hbar}{g\mu_{N}},
\end{equation}
where $\mu_{N}$ is the nuclear magneton, $g$
the g-factor and $\omega_L$ the observable Larmor frequency. 
For interpretation of the physical origin
behind  $B_{hf}$, it is usual to decompose it in four terms:
\begin{equation}
\label{hyperf}
B_{hf}=B_C+B_{orb}+B_{spin}+B_{latt},
\end{equation}
where $B_C$ is the Fermi contact interaction,
of the electronic spin density at
the nucleus with the nuclear magnetic moment, $B_{orb}$ and $B_{dip}$
are the contributions of the magnetic interaction due
to the electronic orbital momentum and electronic 
spin momentum, respectively, and $B_{latt}$ is a 
contribution from the other atomic orbitals in the lattice, usually negligible.

The intrinsic inhomogeneities and remaining distributions 
of point defects lead to the damping of the experimental PAC 
spectrum, which is simulated by a Lorentzian function 
characterized by the $\delta$ (width) parameter.

%%%%%%%%%%%%%%%%%%%%%%%%%%%%%%%%%%%%%%%%%%%%%%%%%%%%%%%%%%%%%%%%%%%%%%%%
%%%%%%%%%%%%%%%%%%%%%%%%%%%%%%%%%%%%%%%%%%%%%%%%%%%%%%%%%%%%%%%%%%%%%%%%
%%%%%%%%%%%%%%%%%%%%%%%%%%%%%%%%%%%%%%%%%%%%%%%%%%%%%%%%%%%%%%%%%%%%%%%%
%%%%%%%%%%%%%%%%%%%%%%%%%%%%%%%%%%%%%%%%%%%%%%%%%%%%%%%%%%%%%%%%%%%%%%%%
%%%%%%%%%%%%%%%%%%%%%%%%%%%%%%%%%%%%%%%%%%%%%%%%%%%%%%%%%%%%%%%%%%%%%%%%

\subsection{Experimental details and sample preparation}
A mixture of radioactive isobars of 
mass $77$, $^{77}$Kr, $^{77}$Br and $^{77}$Se
were produced at the ISOLDE isotope separator online facility at CERN,
and implanted at 30 keV to a dose of approximately $10^{16}$ atoms/m$^2$
 in MnAs samples at room temperature. 
After waiting for the decay
of $^{77}$Kr ($t_{1/2}\approx74$ min.) for $12$ h, the
PAC experiments have started on the decay of $^{77}$Br to $^{77}$Se, after 
annealing, as described. 

One test sample was measured as implanted, 
but the resulting spectrum
was highly attenuated due to implantation defects.
Subsequently, a first annealing step at $600$ C for $200$ s was done in
vacuum ($5\times10^{-4}$ mbar), followed by a fast 
quench to room temperature for both samples, 
after which the spectra substantially improved.

\begin{figure}[ht]\centering
\includegraphics[width=0.5\linewidth]{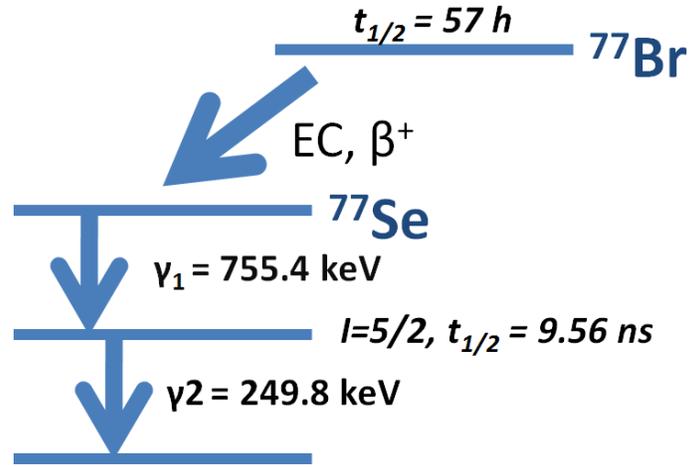}
\caption{\label{cascade}Diagram of the $\gamma-\gamma$ 
decay cascade of $^{77}$Se, with the properties of the 
relevant intermediate isotope, and of the decay from 
the parent isotope $^{77}$Br by processes of electron 
capture and positron emission.}
\end{figure}

The $\gamma-\gamma$
cascade of $^{77}$Br$\rightarrow^{77}$Se is shown in figure~\ref{cascade}. 
The hyperfine interaction is measured in the $9.56$ ns, $249.8$ keV, $I=5/2$ 
intermediate state of the cascade, with quadrupole moment $Q=1.1(5) b$, and 
magnetic moment $\mu=1.12(3)\mu_N$~\cite{Stone2005}.

The directional correlation of the decays is perturbed 
by the hyperfine interactions
and the experimental anisotropy ratio function $R(t)$, 
which contains all of the relevant information,
is expressed as a function of time, $t$. $R(t)=\sum A_{kk}G_{kk}(t)$, 
where $A_{kk}$
are the anisotropy coefficients, depending on the spin 
and multipolarity of
the $\gamma$ decays, and $G_{kk}(t)$ contains the 
information of the hyperfine parameters.
Due to the solid angle attenuation of the detection 
system the anisotropy is reduced, and the effective 
experimental anisotropy was found to be A$_{22}\approx-0.13(1)$.
The long half-life of the parent isotope $^{77}$Br, $57$ h, 
and the relatively short
half-life of the intermediate state, allowed us to 
perform several measurements
with a very good true to chance coincidence ratio 
from a single implantation shot.

The PAC-spectrometer, a high efficiency 
setup of six BaF$_2$ detectors,
provides $30$ coincidence spectra ($6$ 
from $180^{\circ}$\ and $24$ from $90^{\circ}$\ 
between detectors~\cite{BUTZ1989}).

%%%%%%%%%%%%%%%%%%%%%%%%%%%%%%%%%%%%%%%%%%%%%%%%%%%%%%%%%%%%%%%%%%%%%%%%
%%%%%%%%%%%%%%%%%%%%%%%%%%%%%%%%%%%%%%%%%%%%%%%%%%%%%%%%%%%%%%%%%%%%%%%%
\subsection{1st set of PAC experiments}
\label{firstex}
\begin{figure}[ht]\centering
\includegraphics[width=\linewidth]{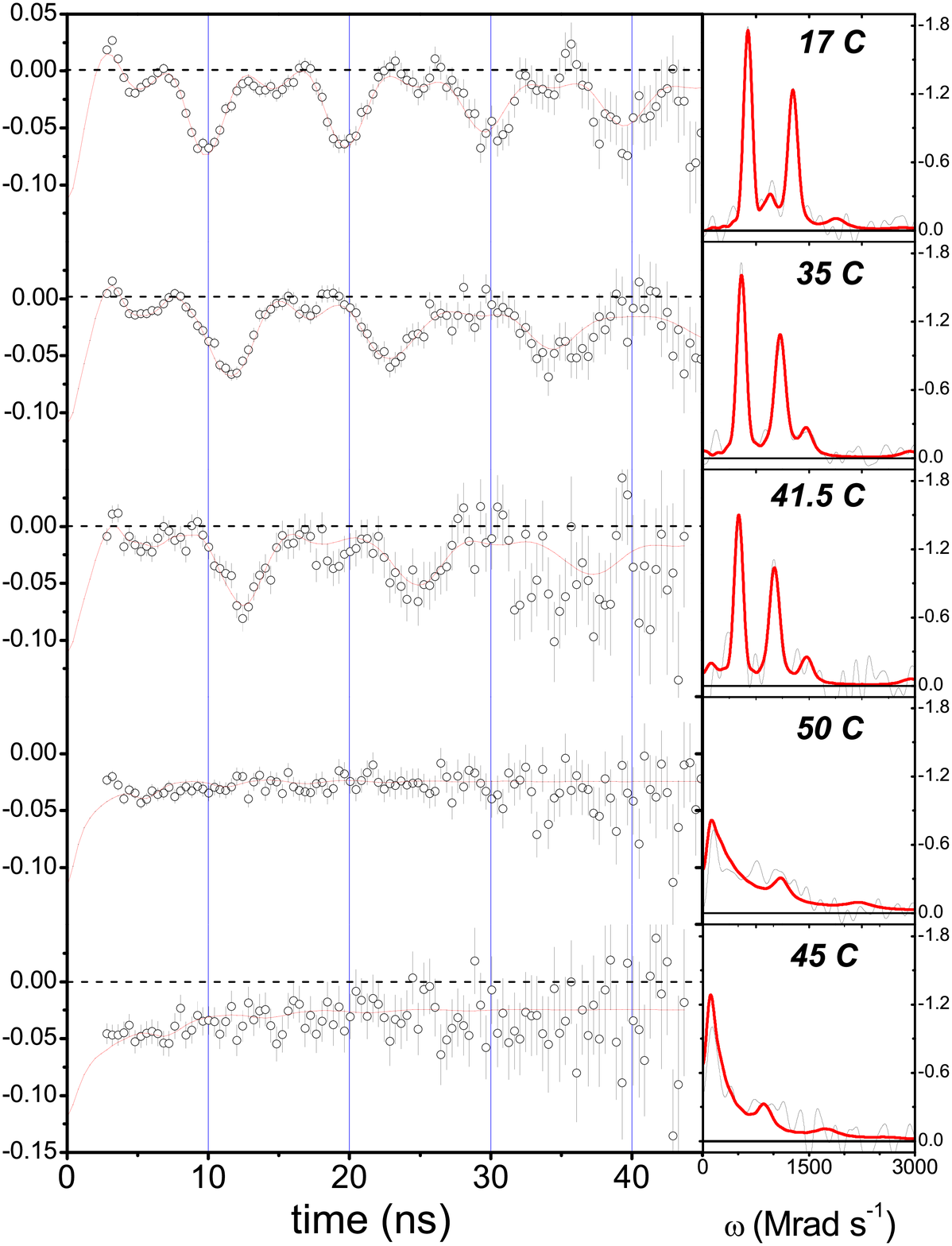}
\caption{(Color online) PAC spectra and Fourier Transforms 
of the first five measurents, in chronological order.
The fits are represented by the red lines.}
\label{PAC_up}
\end{figure}

\begin{figure}[ht]\centering
\includegraphics[width=\linewidth]{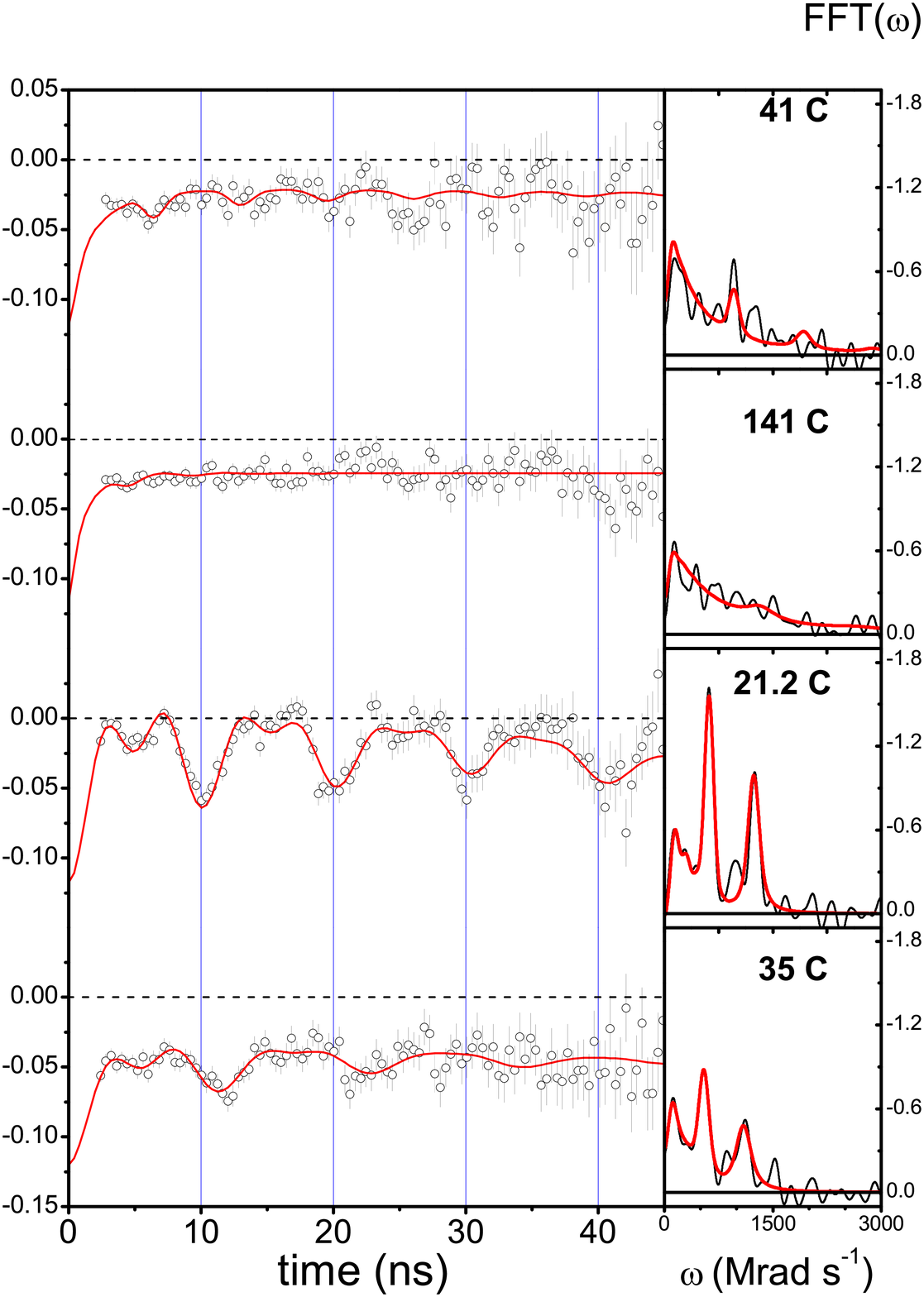}
\caption{(Color online) PAC spectra and Fourier Transforms of the
last five measurents, in chronological order. Between 21.2 and 25
C the sample was heated to 100 C, so that the 35 C measurement is
made on cooling. The fits are represented by the red lines.
}
\label{PAC_dn}
\end{figure}

%%%%%%%%%%%%%%%%%%%%%%%%%%%%%%%%%%%%%%%%%%%%%%%%%%%%%%%%%%%%%%%%%%%%%%%%
%%%%%%%%%%%%%%%%%%%%%%%%%%%%%%%%%%%%%%%%%%%%%%%%%%%%%%%%%%%%%%%%%%%%%%%%
The obtained PAC experimental functions $R(t)$ 
and the respective Fourier analysis are
shown in figures~\ref{PAC_up} and~\ref{PAC_dn}. 

The spectra were fitted using a numerical algorithm
that calculates the hamiltonian of the interaction 
to obtain the magnetic $B_{hf}$ and quadrupole 
EFG parameters~\cite{Barradas1993}. Figs.~\ref{PAC_up} 
and~\ref{PAC_dn} also show the Fourier analysis 
of the $R(t)$ functions for all temperatures measured.

Table~\ref{fitparup} shows all fit parameters 
obtained at the different temperatures, in the 
chronological order of measurements. Between the 
last two measurements, the sample was heated to 
100 C, so the 35 C measurement is made on cooling.

%\begin{table}\centering
%\caption{\label{fitparup}Fit parameters, from measurements when 
%heating and cooling the 1st sample, in chronological order. 
%Fraction (\%) of the measured interactions, quadrupolar 
%frequency ($\omega_{0}$) (Mrad.s$^{-1}$), 
%Larmor frequency $\omega_L$ (Mrad.s$^{-1}$)
%and width of Lorentzian function used in the fitting 
%procedure ($\sigma$) (Mrad.s$^{-1}$). 
%The magnetic frequencies have an estimated upper limit 
%for the error of 15 Mrad.s$^{-1}$. 
%The  fraction $f_3$ and 
%the frequency of the orthorhombic phase were kept 
%fixed at an average value.  H = hexagonal 
%ferromagnetic, O = orthorhombic, H$^*$ = hexagonal paramagnetic.}
%\begin{tabular}{|c|c|c|c|c|c|c|c|c|c|c|}
%\hline\hline
%$T(^{\circ}$C$)$     &
%$f_1$ & $\omega_{L1}$& $\sigma_1$       &
%$f_2$& $\omega_{02}$& $\sigma_2$  &
%$f_3$& $\omega_{03}$&$\sigma_3$& Phase\\
%\hline
%17& 78 & 637 & 21&-&-&-&22& 945& 69  &             \\
%35& 78 & 547 & 28&-&-&-& 22 & 1466  & 69&H\\
%41.5& 78 & 506 & 33&-&-&-& 22 & 1484  & 55&\\
%\hline
%50&-&-&-& 78 & 29& 290& 22& 1108& 10 &\\
%45&-&-&-& 78& 29&  330& 22& 874& 96 &O\\
%41&-&-&-& 78& 29& 290& 22& 963& 46 &\\
%\hline
%141 &-&-&-& 78 & 29&  435& 22& 1324& 17&H$^*$\\
%\hline
%21.2& 78& 618 & 33&-&-&-& 22& 176& 40&H \\
%\hline
%35& 60& 546 & 65&18&28&31& 22& 176& 43& H+O\\
%\hline\hline
%\end{tabular}
%\end{table}

The fit procedure can consider several 
fractions of $^{77}$Se nuclei interacting with different 
hyperfine fields due to different local 
environments. For the ferromagnetic case the fits 
mainly reveal nuclei interacting with a magnetic field. 
 Additionally 
for all phases, a fraction must be 
considered of $^{77}$Se nuclei interacting with 
a strong (EFG$_3$) distribution, 
that we attribute to nuclei on defect regions of 
MnAs which could not be annealed. 
This fraction was firstly allowed to 
vary, but the quality of the fit is not very sensitive to its value. 
In the final fits we constrained this value to 
the average of all previously found values, $f_3=22\%$.
 
Upon the transition the magnetic interaction 
vanishes and a slow frequency, due 
to the orthorhombic phase EFG is revealed. 
The limited time window and low quadrupole moment makes 
it difficult to measure the EFG$_2$ parameters of the 
orthorhombic phase with high precision. Even in the 
most accurate measurements it 
can be fitted reasonably in a large range. On the other hand, 
the stronger EFG$_3$ of f$_3=22\%$ has a large damping which 
also makes its accurate determination difficult. Therefore, 
in the present experiments, the asymmetry 
parameters $\eta_2$, $\eta_3$ were set to zero, since 
large variations produce small changes in the results. On 
the other hand, this procedure agrees with the fact that 
the hexagonal and weakly distorted orthorhombic 
symmetries produce very small axial asymmetry parameters. 
The frequency $\omega_{02}$
was also fixed in an average value.

%On 
%the following we detail the results, 
%quantifying the fraction of $^{77}$Se nuclei, 
%interacting with different 
%fields as a function of temperature.

The magnetic phase, characterized by a well 
defined magnetic hyperfine field, could be characterized also by a very small 
EFG. The fit program properly handles this problem by resolving the 
Hamiltonian for the combined interaction.
%This simultaneous presence of magnetic and quadrupole 
%EFG fields gives rise to ``combined interactions''. The 
%fit program properly handles this problem by resolving 
%the Hamiltonian for the combined interaction. 
In the results we present only a 
pure magnetic interaction, since with a combined interaction the 
EFG (V$_{zz}$) would have to be very small in this phase and cannot 
be properly disentangled within the short analysis 
time of $45$ ns. We estimate a majorant 
for $V_{zz}\lesssim1\times 10^{21}V/$m$^2$, above 
which the quality of the fit would significantly 
degrade. 

 The obtained V$_{zz}$ attributed to the orthorhombic 
phase is less than $1.1\times 10^{21}V/$m$^2$ at all temperatures measured. 

The frequencies $\omega$ and Lorentzian widths $\sigma$ 
are similar for the whole temperature range in this 
phase. However, we point that before the experiment 
performed at 141 C during six hours, EFG$_2$ shows a 
relevant attenuation of $\sigma_2\approx300$ Mrads$^{-1}$. After 
this measurement the attenuation was considerably reduced 
and the characteristic EFG$_3$ parameters attributed to Se 
interacting with defects of MnAs have considerably changed. Both 
modifications compare well with what is observed in the second 
set of PAC experiments immediately after the 600 C annealing step. We 
think this is evidence for an incomplete annealing that 
was compensated during the lenghty six hours measurement at 141 C.
 
The EFG\ parameters measured at 141 C, above the second-order
 phase transition,
shows a very low V$_{zz}$ as expected from the NiAs-type 
structure, and there is no hyperfine field since the 
sample should be paramagnetic, following a Curie-Weiss law at this temperature. 

The measurement at 35 C shows a lower amplitude of the R(t) 
function due to the coexistence of hexagonal and orthorhombic 
phases. Still, there is a stronger attenuation of the 
magnetic field that can correlate with the dynamics of the phase coexistence.

The first-order transition reported in the literature  
when heating is clearly seen in the PAC
 spectra at approximately $42$ C with the disappearing 
magnetic hyperfine field when measuring at $50$ C.

The fact that the spectrum measured at $41.5$ C (when 
raising T, fig.~\ref{PAC_up})
and the spectrum measured at $41$ C (when lowering T, fig.~\ref{PAC_dn} )
are markedly different, shows that the hysteretic 
behavior of the macroscopic magnetization usually 
measured is also present at the microscopic-local like hyperfine field.

%%%%%%%%%%%%%%%%%%%%%%%%%%%%%%%%%%%%%%%%%%%%%%%%%%%%%%%%%%%%%%%%%%%%%%%%%%%%%%%%
%%%%%%%%%%%%%%%%%%%%%%%%%%%%%%%%%%%%%%%%%%%%%%%%%%%%%%%%%%%%%%%%%%%%%%%%%%%%%%%%
%%%%%%%%%%%%%%%%%%%%%%%%%%%%%%%%%%%%%%%%%%%%%%%%%%%%%%%%%%%%%%%%%%%%%%%%%%%%%%%%
%%%%%%%%%%%%%%%%%%%%%%%%%%%%%%%%%%%%%%%%%%%%%%%%%%%%%%%%%%%%%%%%%%%%%%%%%%%%%%%%

\subsection{2nd set of PAC experiments - First-Order Transition}
\label{secondex}

A detailed study of the first-order transition has 
been done according the following order, on a second 
sample: 21.1, 40.8, 41.3, 42.3, 43.5, 124.5 (raising temperature); 
41.4, 39.5, 37.5, 36.6, 33.3, 32.5, 29.9, 13.3, -196 C (lowering temperature).

\begin{figure}
\includegraphics[width=\linewidth]{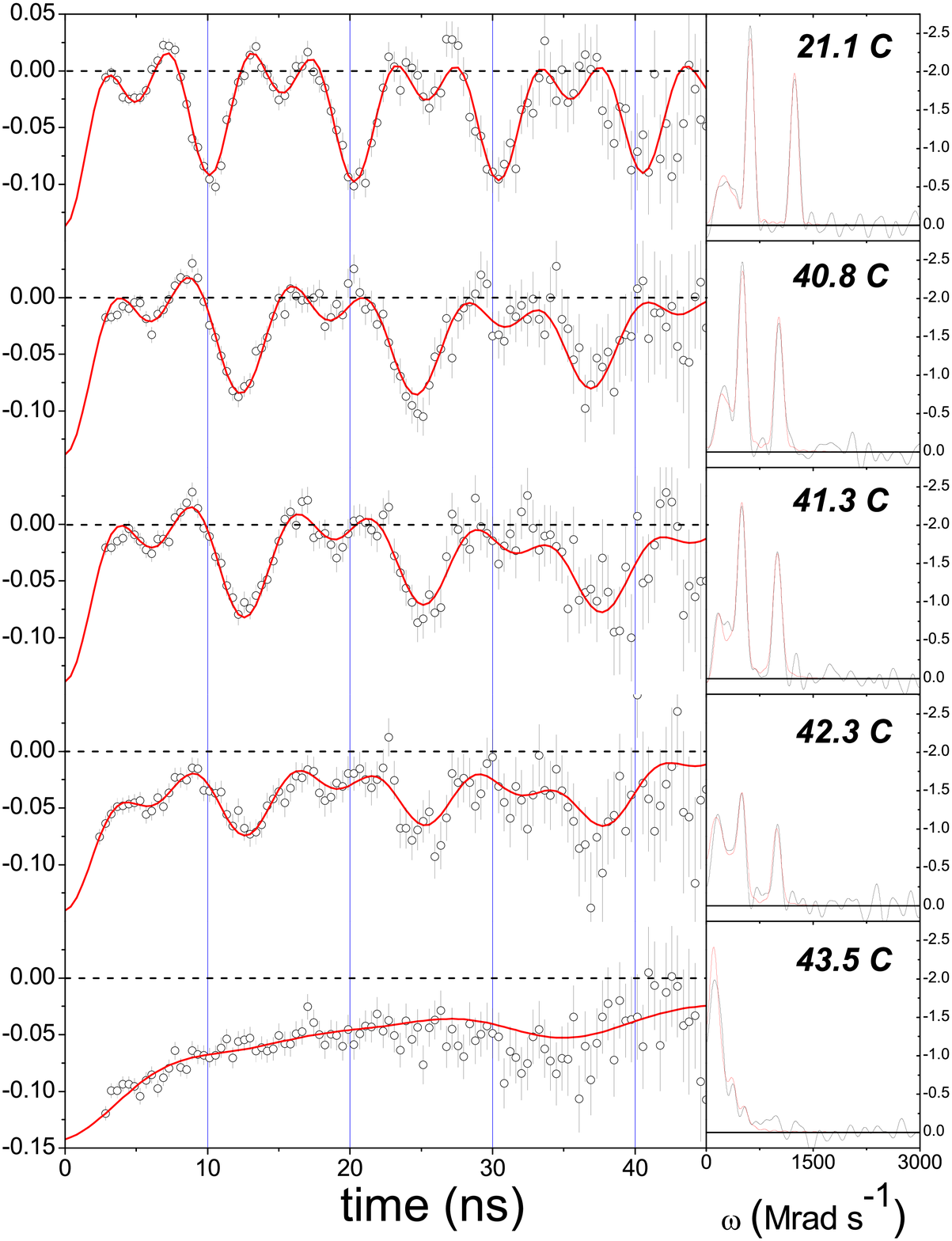}
\caption{\label{fig:rt1}(Color online) PAC spectra and 
Fourier Transforms. Measurements made when heating the sample.
The fits are represented by the red lines.}
\end{figure}

The first five PAC measurements, done when heating the 
sample, from room temperature, to above the first-order 
transition, are shown in figure \ref{fig:rt1}.  

\begin{figure}
\includegraphics[width=\linewidth]{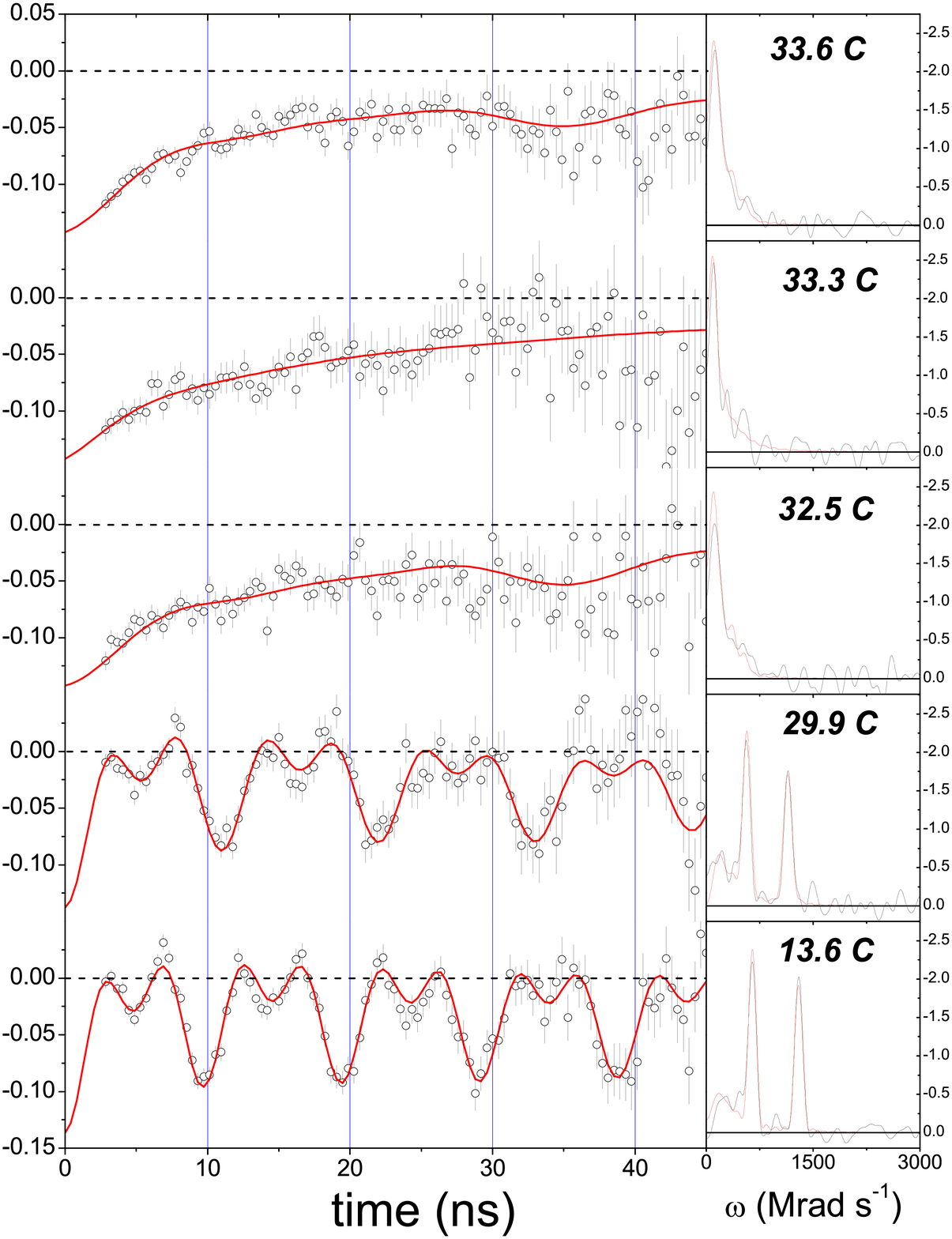}
\caption{\label{fig:rt3}(Color online) PAC spectra 
and Fourier Transforms. Measurements made when cooling the sample.
The fits are represented by the red lines.}
\end{figure}

The five  measurements, done when cooling the 
sample, from 36.6 to 13.3 C, also passing the 
transition, are shown in figure \ref{fig:rt3}. 

The measurements done when cooling the sample 
above the phase transition (41.1, 39.5, 37.5, 36.5 C), coming 
from a high temperature (124.5 C), are not included, since 
those spectra are similar to the first spectrum in figure \ref{fig:rt3}, 
at 36.6 C.

\begin{figure}
\includegraphics[width=\linewidth]{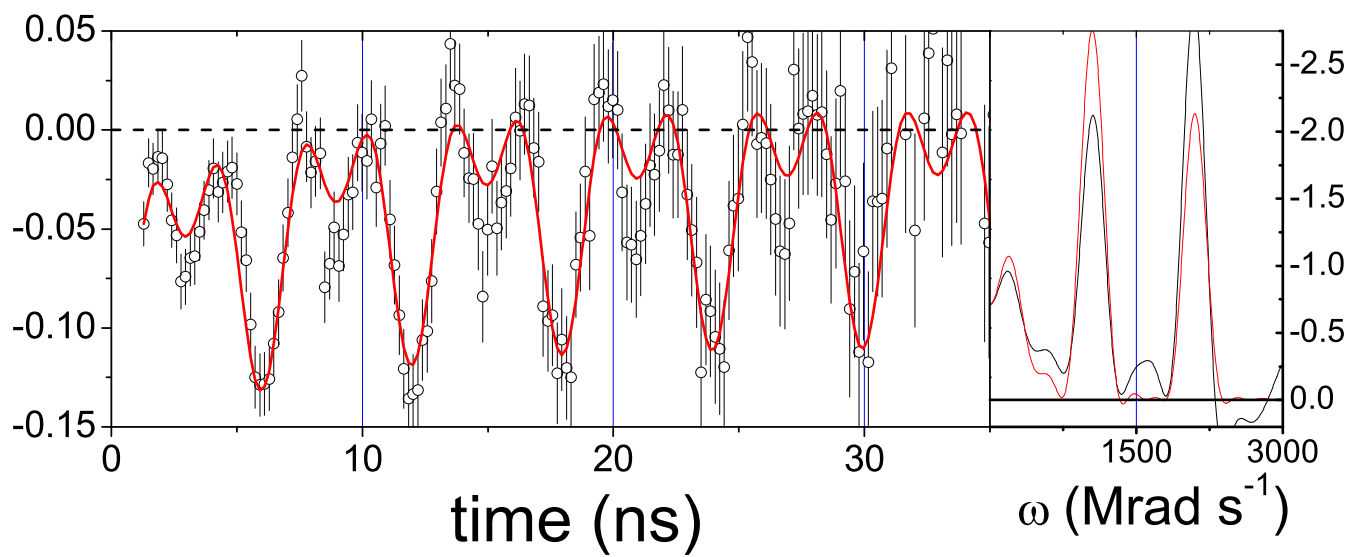}
\caption{\label{fig:lno} (Color online) PAC spectra 
and Fourier Transforms at -196 C.
The fits are represented by the red lines.}
\end{figure}

A last measurement performed  with the sample at liquid 
nitrogen temperature is shown in figure~\ref{fig:lno}.

In a similar way to the preceding section, the fits were 
done considering a magnetic hyperfine field and low EFG, which 
are characteristic of each phase. For the reasons already 
detailed the asymmetry parameter is set to zero for all EFGs, 
and the fraction attributed 
to defect and orthorhombic frequencies are fixed in average values.
Also, an additional EFG characterized by a quadrupole 
interaction of $\omega_0\approx176$ 
Mrad.s$^{-1}$, V$_{zz}\approx 6.4 \times 10^{21}$ V/m$^2$, is 
found that accounts for 30\% of the probe nuclei in 
perturbed environments of the sample, still remaining after annealing. 

The values of all fitted parameters can be found in table~\ref{fitpar2}.

As can be seen by the changes in the spectra, 
the transformations occur near $T_{C,i} \approx 42.3-43.5 C$ 
and $T_{C,d}=30-32.6$ C. Therefore we estimate the thermal irreversibility 
to be between 10-13.5 C. This hysteresis is somewhat larger
than that reported in other works by X-ray and 
magnetization measurements (10 C)
~\cite{MnAs:Goo67,Nascimento2006,MnAs:Wil64,JJAP.42.L918,Ishikawa2004408}.

\begin{figure}
\includegraphics[width=\linewidth]{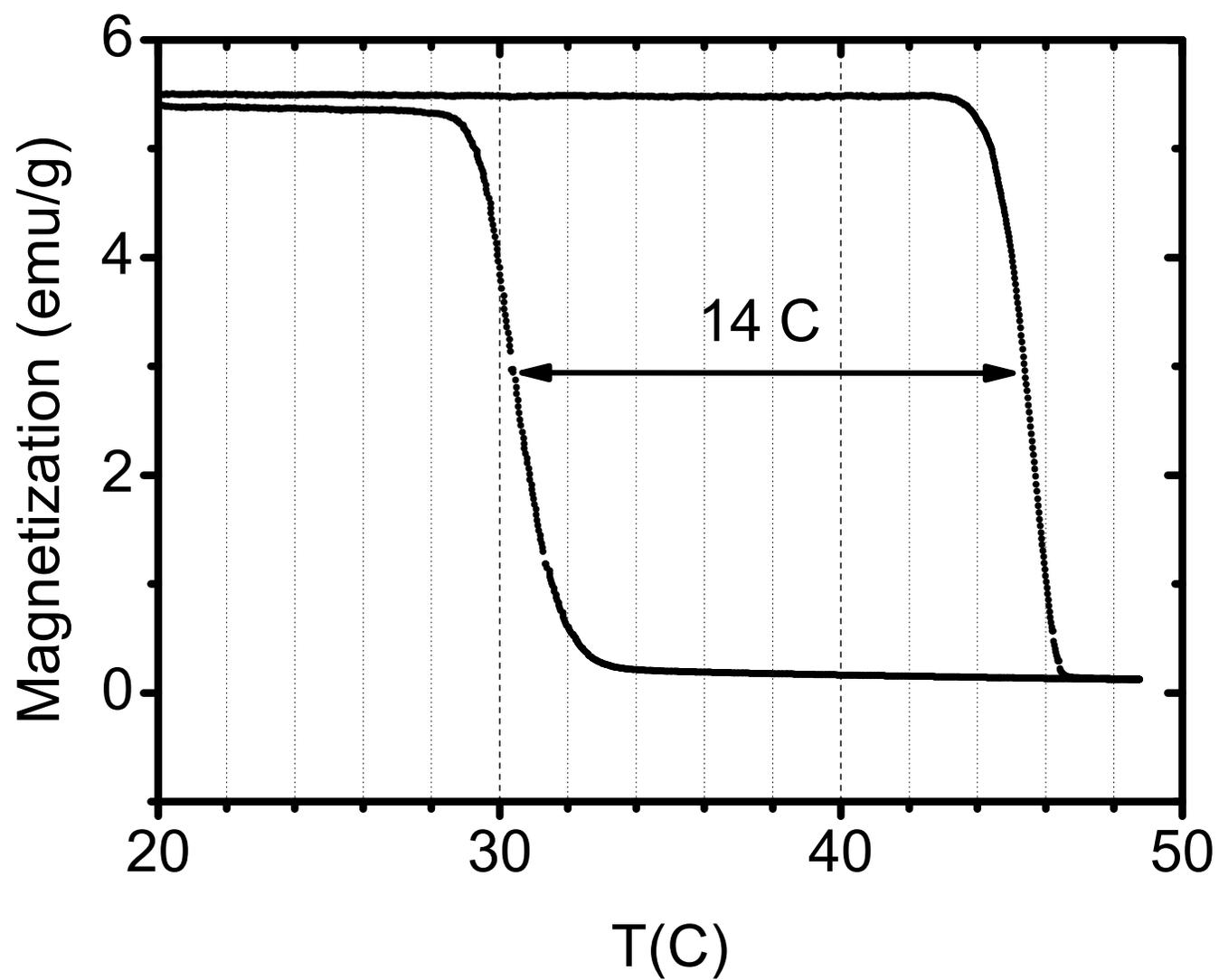}
\caption{\label{mag} Temperature dependence of the 
magnetization, measured at $B=0.01$ T, signalling 
the strong thermal hysteresis at the 1st order transition.}
\end{figure}

Figure \ref{mag} shows magnetization measurements on the same samples
with a vibrating sample magnetometer 
with $B=0.01$ T, showing T$_{C,i} = 45$ C 
and T$_{C,d} = 30.7$ C. The abrupt change over 
2 C, at $\approx 44$ C, when heating (see fig.~\ref{mag}), is 
in agreement with the hyperfine field changes measured above $43.5$ C.

As in the first set of PAC measurements, there is a 
hyperfine field of 24 T just before the transition. Then 
the magnetic phase disappears in a small temperature interval, as 
shown in the spectra of figure~\ref{fig:rt1} at $42.5$ and $44.6$ C. This 
shows no continuous decrease of the hyperfine field to 
zero before the transition. Note that this conclusion 
cannot be learned from macroscopic magnetization measurements 
only (see figure \ref{mag}), where the magnetization 
can be seen to decrease to zero, since just before the 
ferromagnetic to paramagnetic phase transition and within the temperature 
difference of ~1 C, only a very small 
variation of the hyperfine field is observed.  This clearly shows that the 
magnetization changes are mainly the result of changes in phase fractions, 
instead of thermal disorder.%, in agreement 
%with what was found in epitaxial thin films~\cite{Adas}. 

\begin{figure}
\includegraphics[width=\linewidth]{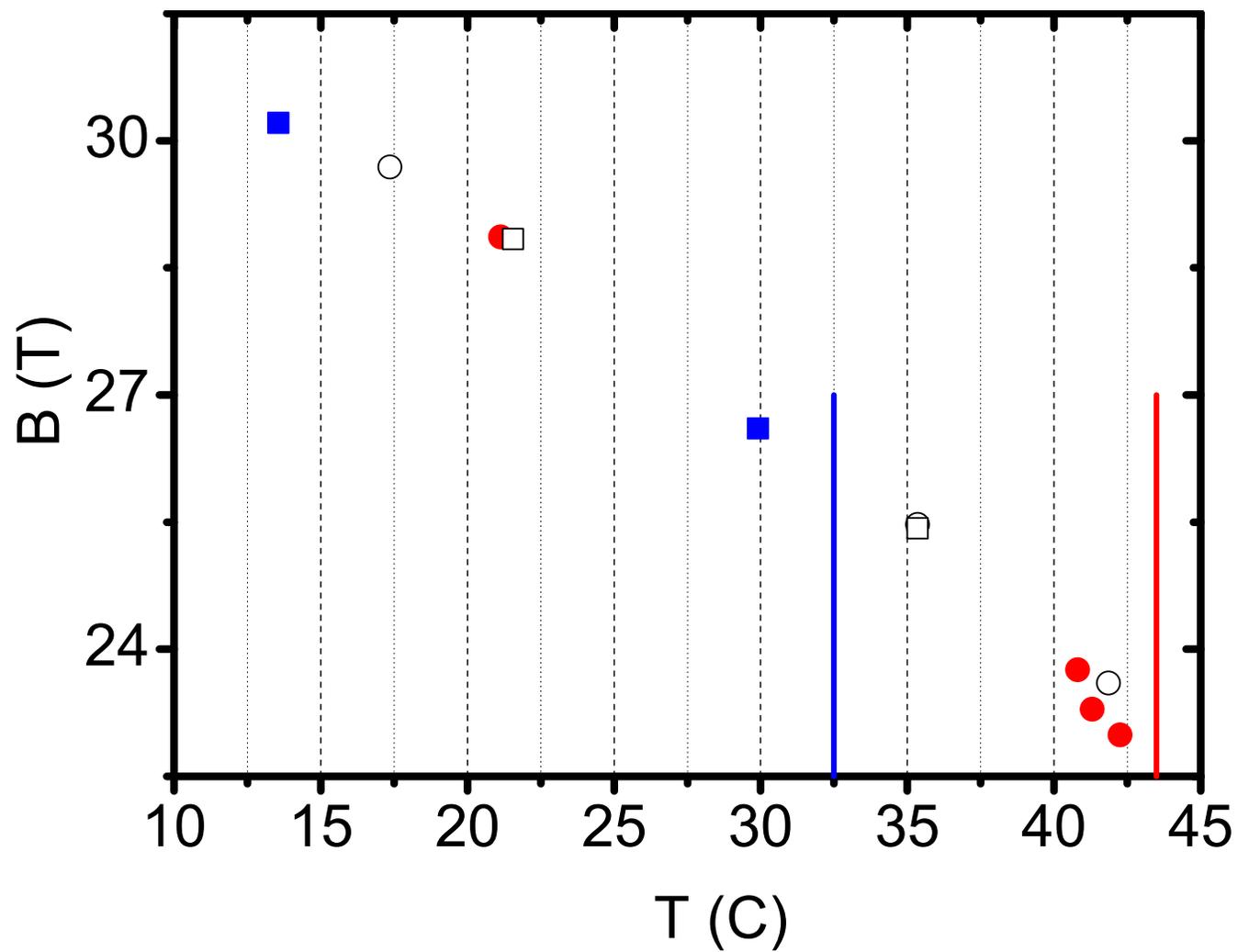}
\caption{\label{asda}(Color online) Hyperfine field 
of the main fraction vs. temperatures, excluding 
the value at -196 C. Circles for measurements when 
heating, squares for cooling.  The open symbols 
show the results of the first experiment, for comparison. The 
lines are the first (when heating) and last (when cooling) 
measurements at the orthorhombic phase.}
\end{figure}

Figure \ref{asda} shows the hyperfine fields 
obtained, where the agreement for both 
experiments can clearly be seen.
 
Notice that the attenuation of the hyperfine fields in 
the ferromagnetic phase increases towards the phase 
transition temperature. When cooling from high temperature 
the same behavior is observed and at liquid nitrogen 
temperature (- 196 C), no attenuation is observed. These 
observations hint at dynamic processes due to spin fluctuations.
 
At $42$ C the amplitude of the magnetic part of the R(t) 
functions is smaller than at lower temperatures, showing 
a reduced fraction of the ferromagnetic phase still present. A 
fraction of 60\% for the $^{77}$Se atoms at the ferromagnetic phase 
while other 10\% show a small quadrupole frequency.
 This third fraction has V$_{zz}= 1.08$ ($\eta$ was fixed to zero), 
corresponding
to the value found for the other fits in the orthorhombic phase. 
We can say that the phase 
coexistence only occurs in a width of 2.2 C or less, since 
the measurement 
below and at 
($41.3$ C) and above and at ($43.5$ C) show only the 
ferromagnetic and paramagnetic phases, respectively. This width is 
in agreement with previous measurements of 
approximately $2$ C~\cite{MnAs:Mir2003,Gama2009}.

The 1st order structure transformation was also probed 
with temperature dependent X-ray powder diffraction studies 
in a Philips diffractometer. We performed detailed measurements as a function 
of temperature, in three selected diffraction angle regions, 
were changes in the transition are easily seen.
Three $2\theta$ intervals were 
selected, 31.4-32.6, 41.8-43.1, and 48.6-50 degrees, where 
one peak characteristic of the hexagonal phase disappears 
in the transition with the appearence of peaks 
characteristic of the orthorhombic phase. In the 
31.4-32.6 $2\theta$ interval the (101) peak disappears 
with the appeareance of (102) and (111) peaks 
almost at the same angle. The same situations occur 
when the (102) peak disappears and (202), (211) peaks 
appears in the orthorhombic phase for 41.8-43.1 
degrees. For 48.6-50 degrees, the (110) hexagonal 
peak transforms into (013), (020), (212) and (301) peaks 
in the orthorhombic phase.
The fit of the peaks was done simply with gaussian 
functions, one gaussian for the hexagonal peak and 
another gaussian for the two or more highly overlapped 
orthorhombic peaks. The areas of each peak should 
correspond approximately to the fraction of each phase. 

\begin{figure}
\includegraphics[width=\linewidth]{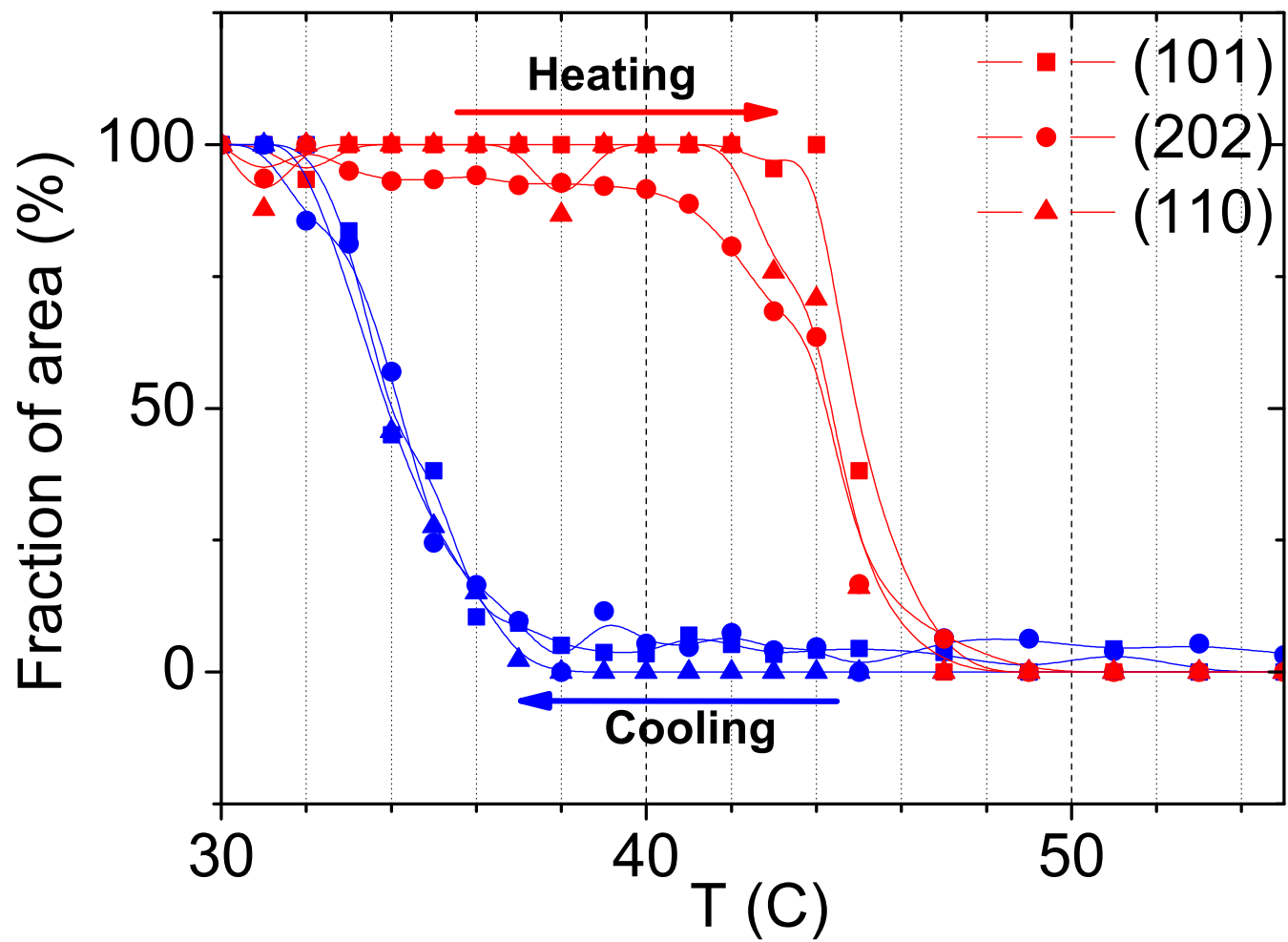}
\caption{\label{fracX}(Color online) XRD Fractions of the 
total area corresponding to the peaks of the hexagonal 
phase, at three $2\theta$ intervals. }
\end{figure}

Figure \ref{fracX} shows the fraction of the 
hexagonal phase obtained this way for the 3 angular intervals. 
The hysteresis produces here a difference 
of approximately 12 C. Magnetization measurements 
show a somewhat higher thermal hysteresis difference at half 
height, 14 C, which 
might indicate that the magnetic coupling is 
disturbed before the hysteresis is completed. 
The thermal hysteresis interval obtained from PAC has a large uncertainty (10-13.5 C) 
 but is in agreement with both measurements.

%%%%%%%%%%%%%%%%%%%%%%%%%%%%%%%%%%%%%%%%%%%%%%%%%%%%%%%%%%%%%%%%%%%%%%%%%%%%%%%%
%%%%%%%%%%%%%%%%%%%%%%%%%%%%%%%%%%%%%%%%%%%%%%%%%%%%%%%%%%%%%%%%%%%%%%%%%%%%%%%%
%%%%%%%%%%%%%%%%%%%%%%%%%%%%%%%%%%%%%%%%%%%%%%%%%%%%%%%%%%%%%%%%%%%%%%%%%%%%%%%%
\section{First-principles calculations}

Knowing the lattice location of the PAC probe is 
of fundamental importance to understand the values obtained.
After the decay of $^{77}$Br, it is expected that 
the $^{77}$Se PAC probe may be substitutional at the 
As site, since As has a similar atomic radius and a 
neighbor atomic number. In order to check this assumption 
and to see the differences in the hyperfine
 parameters actually measured at the probe site, we 
have used ab-initio density functional calculations.

Despite the existence of some published works reporting 
ab-initio simulations in this 
system, the hyperfine parameters are usually not reported.
The work of Ravindran et al.~\cite{MnAs:Rav99},
presents the calculation of the hyperfine parameters and 
magneto-optical properties,
 using density functional calculations, of three 
manganese pnictides MnX (with X=As, Sb and Bi).
 Their calculations used the FLAPW method and the 
electric field gradient and
magnetic hyperfine field were presented. However, 
as suggested by A.~Svane~\cite{MnAs:Sva2003},
the calculated structures of~\cite{MnAs:Rav99} 
are incorrect, since the positions of Mn and the pnictide were 
exchanged with respect to the the stable NiAs-type structure. 
%Consequently the calculations were
%made with the unobserved anti-NiAs 
%structure (with the atomic positions of Mn and As exchanged),
%but were still compared with experiments.
%We have calculated the anti-NiAs structure
%for MnAs and reproduced the results of Ravindran et al.,
%thus confirming the suggestion from~\cite{MnAs:Sva2003}.
Recently, calculations of the hyperfine 
parameters in bulk and surfaces of MnAs were also reported, 
and the correct values were obtained~\cite{jamal}. 

Here we also calculate the hyperfine parameters with the similar
full potential (L)APW+lo
method, as implemented in the \textsc{Wien2k} code~\cite{wien2k}. In this 
method the space is
divided in spheres, centered at the atoms, 
where the valence states are described by 
atomic-like functions, and the interstitial space, where plane waves 
are used.

The Mn and As atomic spheres used have both a radius
of $2.5$ a.u.\ . We checked convergency of the hyperfine 
parameters and total energy as a function of the number 
of k-points used for integration in the Brillouin zone and 
the number of plane waves in the basis.
The calculations are spin-polarized and consider a
ferromagnetic arrangement of Mn moments.
For the calculations of hyperfine parameters, spin-orbit coupling 
is included, for the other properties no spin-orbit coupling is 
included, with a scalar-relativistic basis for the valence 
electrons, while for the  core electrons the treatment is 
always fully-relativistic. For Mn core states 
are 1s, 2s, 2p, and 3s, and valence states 
are 3p, 3d, and 4s, while for As 1s, 2s, 2p, 3s, and 3p are 
core states and 3d, 4s, and 4p are valence states.
The PBE Generalized Gradient Approximation~\cite{PhysRevLett.77.3865} 
exchange-correlation
functional is used, since the LSDA is known to 
give poor results in
this compound~\cite{PhysRevB.65.113202}.

The calculated EFG of the MnAs sites in the hexagonal 
phase is shown in the table~\ref{EFGAs}. Due to the hexagonal
symmetry $\eta$ is zero, and the direction of the 
principal axis of the EFG tensor
 is parallel to the c-axis.
The EFG inside the spheres, which is almost equal 
to the total EFG,
can be separated in different contributions, since 
the the states are described in combinations of spherical harmonics,
 with different angular momentum components. In 
this case the p-p and d-d contributions of the density 
are the dominant terms, with 
V$_{zz}^{pp}\propto\langle1/r^3\rangle_p[1/2(p_x+p_y)-p_z]$ 
and V$_{zz}^{dd}\propto\langle1/r^3\rangle_d[(d_{xy}+d_{x^2-y^2})-1/2(d_{xz}+d_{yz})-d_{z^2}]$.
For Mn, V$_{zz}^{pp} = -1.45$ and V$_{zz}^{dd} = -1.70$  $\times 10^{21}$V/m$^2$,
states with both p and d character contribute to the total EFG.
For the As atoms, the states of p character are the
dominant contribution with V$_{zz}^{pp} = 1.27$ 
and V$_{zz}^{dd} = 0.06 \times 10^{21}$ V/m$^2$.
Cutting the 3d$^{10}$ states out of the density 
calculation the V$_{zz}$ at As remains almost the same,
confirming that the contribution from the As 
filled d electrons is negligible.

In order to improve the 
results of the previously mentioned work,
we also calculated the EFG at MnSb and MnBi.
We discuss them in section~\ref{pnictides},
along with other quantities.

The Fermi contact hyperfine field at 
the nucleus is calculated, with
the electron density averaged at a sphere with the
Thomson radius, $r_T=Ze^2/mc^2$, according to the
formulation of Bl\"{u}gel et al.~\cite{blugel} in
which $\overrightarrow{B}_C=\frac{8\pi}{3}\mu_B\overrightarrow{m}_{av}$, i.\ e. 
the contact hyperfine field is parallel to the average spin density.
The contributions of the contact hyperfine field due
to core and valence electron density contributions
are discriminated in the tables. We remark the 
fact that while in As the hyperfine field is
 determined almost exclusively by its valence 
contribution, caused by the polarization by Mn atoms, 
 the core and valence contributions of Mn cancel in a large amount. 
This is due to the 
core polarization mechanism~\cite{Blugel1987}, where the
 core hyperfine field in Mn has a
negative sign due to the polarization of core $s$ electrons by the $d$ shell: 
the majority electrons are
attracted to the polarizing $d$ electrons while the minority electrons are 
repelled, resulting in an excess minority charge at the nucleus.
The on-site orbital and spin dipolar 
contributions are also calculated. These contributions
are small when compared with the contact hyperfine field. 
In order to see the change due to different lattice parameters
in the hyperfine fields, we calculated also 
with the low temperature
lattice constants~\cite{JPSJ.51.3149}, and the 
obtained values are almost equal (tables~\ref{BHF1} and~\ref{BHF2}). This simply 
shows that the collinear spin density functional theory 
calculations cannot reproduce temperature related 
changes based only on the lattice constants. 

The previously obtained hyperfine parameters are in reasonable 
agreement with the GGA calculations of Jamal et al.~\cite{jamal}. 
Some differences are expected, since while their  calculations consider the 
full theoretical lattice optimization, we only minimized the atomic forces 
keeping the lattice parameters fixed at the experimental values.
Relative to their results, for the V$_{zz}$ at Mn and As,
small differences of 5\%  (1.53 against 1.46) and 4\% (-3.63 against -3.78) 
are obtained, respectively. 
For the hyperfine fields the differences are -9 (present work) 
compared with 1 T at the As site 
(small absolute difference) , and 24.7 (present work) 
compared with 31.8 T at the Mn and As sites.

To compare with the PAC results using the 
implanted probe, the presence of a highly diluted (ppm) Se probe 
must be accounted for in supercell calculations.
The EFG and hyperfine field were calculated for hypothetical
situations where the Se is substituted at As 
and at Mn sites using Mn$_{15/16}$Se$_{1/16}$As
and MnSe$_{1/16}$As$_{15/16}$) supercells.

The results for supercells with Se concentration 
of $1/16$ are shown in the table~\ref{EFGSe}.
The atomic forces were not high, and were 
minimized by moving the free atomic coordinates.
The small changes in this type of system due
to the lattice constants (tables~\ref{BHF1} and~\ref{BHF2})
motivated us to keep using the MnAs room 
temperature lattice constants.
The hyperfine field calculated with the Se 
atom substitutional at the As or Mn sites 
would be exact only at 0K (disregarding zero-point effects, 
which should be small~\cite{Torumba2006}). 
Our closest measured value is
at liquid nitrogen.

There is a good agreement of the measured 49 T at 77K 
when compared with the calculated 54.3 T at the As site.
In contrast, for the case in which Se is substitutional 
at the Mn site, $|$B$_{hf}|$ is too low when compared to 
the experiment, even near the transition, and the 
very high V$_{zz}=17.4 \times 10^{21}$ V/m$^2$ immediately
discards the possibility that the probe is located there, whereas
the EFG is very small for As-site substitutional 
Se, in agreement with experiment.
The calculation of the formation energies $\Delta H_f$ for 
the two substitutions also indicates this assignment,

\begin{equation}
\Delta H_f=E^{sup}_{imp}-8\times E^{MnAs}-\mu_{Se}+\mu_{As/Mn}
\end{equation}
 where $E^{sup}_{imp}$ is the total energy of 
the  $2\times2\times2$ supercell with a Se impurity, $E^{MnAs}$ is 
the energy calculated for the pure compound, and $\mu_{Se}$ is taken 
as the total energy of nonmagnetic $hcp$ Se. The chemical potential 
of As or Mn (according to the substituted site) is set as the energies 
of fcc antiferromagnetic Mn and nonmagnetic rhombohedral As. The formation energy 
obtained for substitution at the As site is 0.03 eV, while for the Mn 
substitution it has a higher value of 
2.84 eV, confirming the hyperfine calculation. 
However, since Br is the implanted atom ,
if there is no time for relocation between the 
Br$\rightarrow$Se decay and the PAC measurement, the 
formation energy of Br should be a better indication.
Therefore, we also calculate these formation energies, 
using the energies of supercells 
of the same size for the same substitutions, with Br, 
and the energy of nonmagnetic solid Br$_2$ as $\mu_{Br}$, 
instead of $\mu_{Se}$, in the previous formula.
 The obtained results are 0.94 eV for the Br at As substitution, 
and 3.95 eV at the Mn site, again confirming the As substitution.

%%%%%%%%%%%%%%%%%%%%%%%%%%%%%%%%%%%%%%%%%%%%%%%%%%%%%%%%%%%%%%%%%%%%%%%
%%%%%%%%%%%%%%%%%%%%%%%%%%%%%%%%%%%%%%%%%%%%%%%%%%%%%%%%%%%%%%%%%%%%%%%
\section{Manganese Pnictides}
\label{pnictides}
%%%%%%%%%%%%%%%%%%%%%%%%%%%%%%%%%%%%%%%%%%%%%%%%%%%%%%%%%%%%%%%%%%%%%%%
%%%%%%%%%%%%%%%%%%%%%%%%%%%%%%%%%%%%%%%%%%%%%%%%%%%%%%%%%%%%%%%%%%%%%%%

Full potential calculations of the hyperfine parameters and
other properties of manganese pnictides 
were performed by Ravindran et al.~\cite{MnAs:Rav99},
but with the anti-NiAs structure.
Here we report the same properties as 
calculated with the FLAPW method, i.~e. the spin magnetic moments,
the density of states and the hyperfine parameters,
 with the NiAs-type structure.

The hyperfine parameters are especially 
sensitive to the type of structure. 
For the atoms of Mn and As, in the true 
structure  V$_{zz}=-3.7$ and $1.4\times10^{21}$V/m$^2$, respectively, while 
in the anti-MnAs structure  V$_{zz}=0.4$ and $11.8\times10^{21}$V/m$^2$.

Table~\ref{EFGpnictides} shows the EFG of the three manganese pnictides.
The asymmetry parameter and V$_{zz}$ direction are omitted,
since they are always 0 and (0,0,1).
The EFG of the pnictide site increases with 
increasing atomic number (As, Sb, Bi), which coincidentally 
also happens in the work of
Ravindran et al.~\cite{MnAs:Rav99}.
For the EFG of Mn the situation is reversed,
in our calculations its absolute value increases,
while their results with the inverted structure 
have a slight decrease ascribed to volume effects, which cannot be true now.

The spin moments for each atom, calculated inside the LAPW spheres, are
presented in table \ref{MM}.
Experimental values and values obtained from other
band-structure calculations are also presented.
With our calculation, the values obtained, in $\mu_N$ per formula unit,
are now in a better agreement with experiment.
Similar calculations (references in table \ref{MM}), which 
have used the NiAs-structure,
get values which are in accordance to our results.
consistently lower than experiment. 

\begin{figure}
\includegraphics[width=\linewidth]{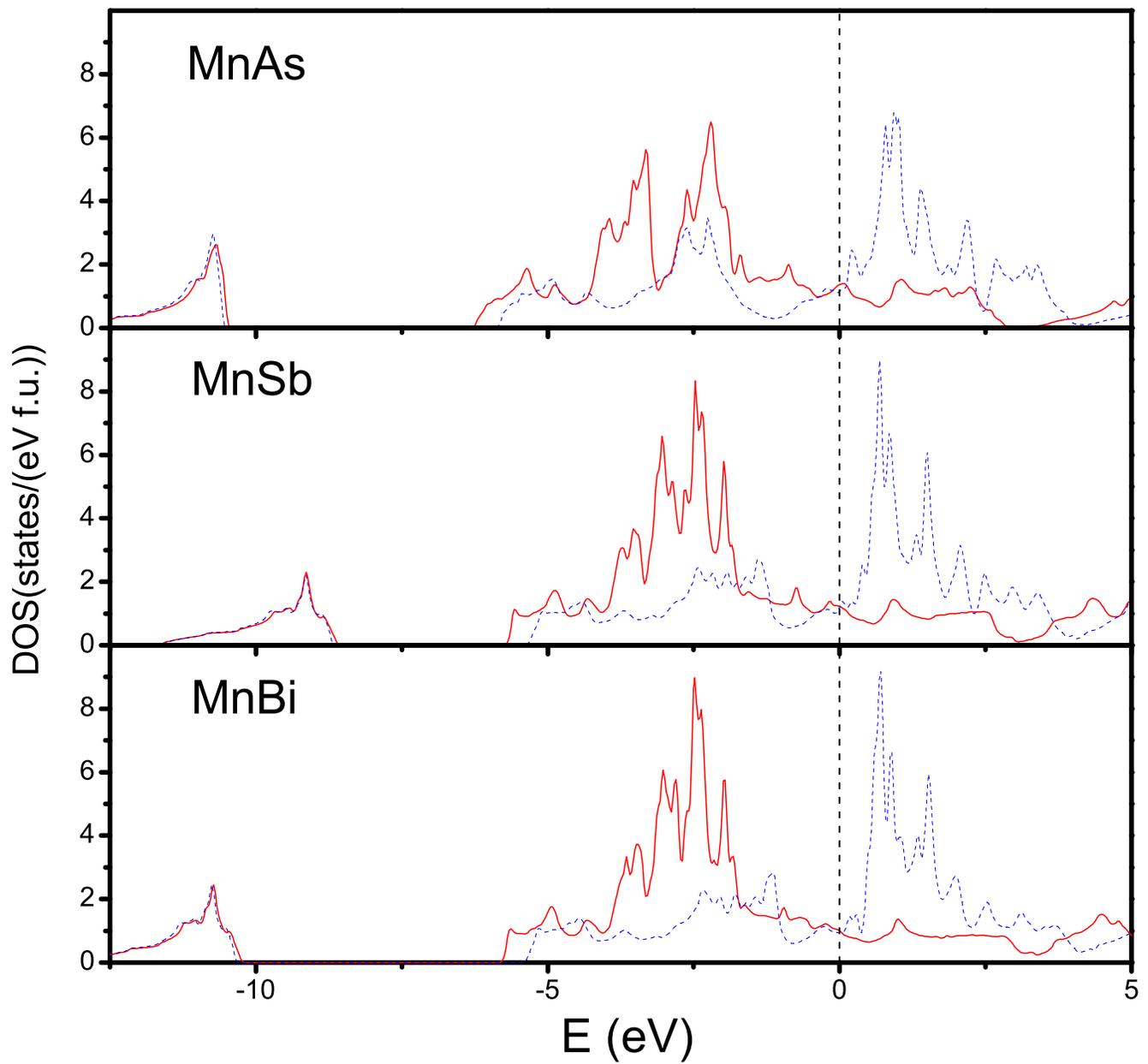}
\caption[]{(Color online) Total density of states for 
MnAs, MnSb, and MnBi in units of $states/(eV f.\ u.)$. The up 
states are represented by the red full line and the down 
states by the blue dashed line. Energy is in eV, relative 
to the Fermi energy (dashed line).}
\label{DOS}
\end{figure}

\begin{figure}
\includegraphics[width=\linewidth]{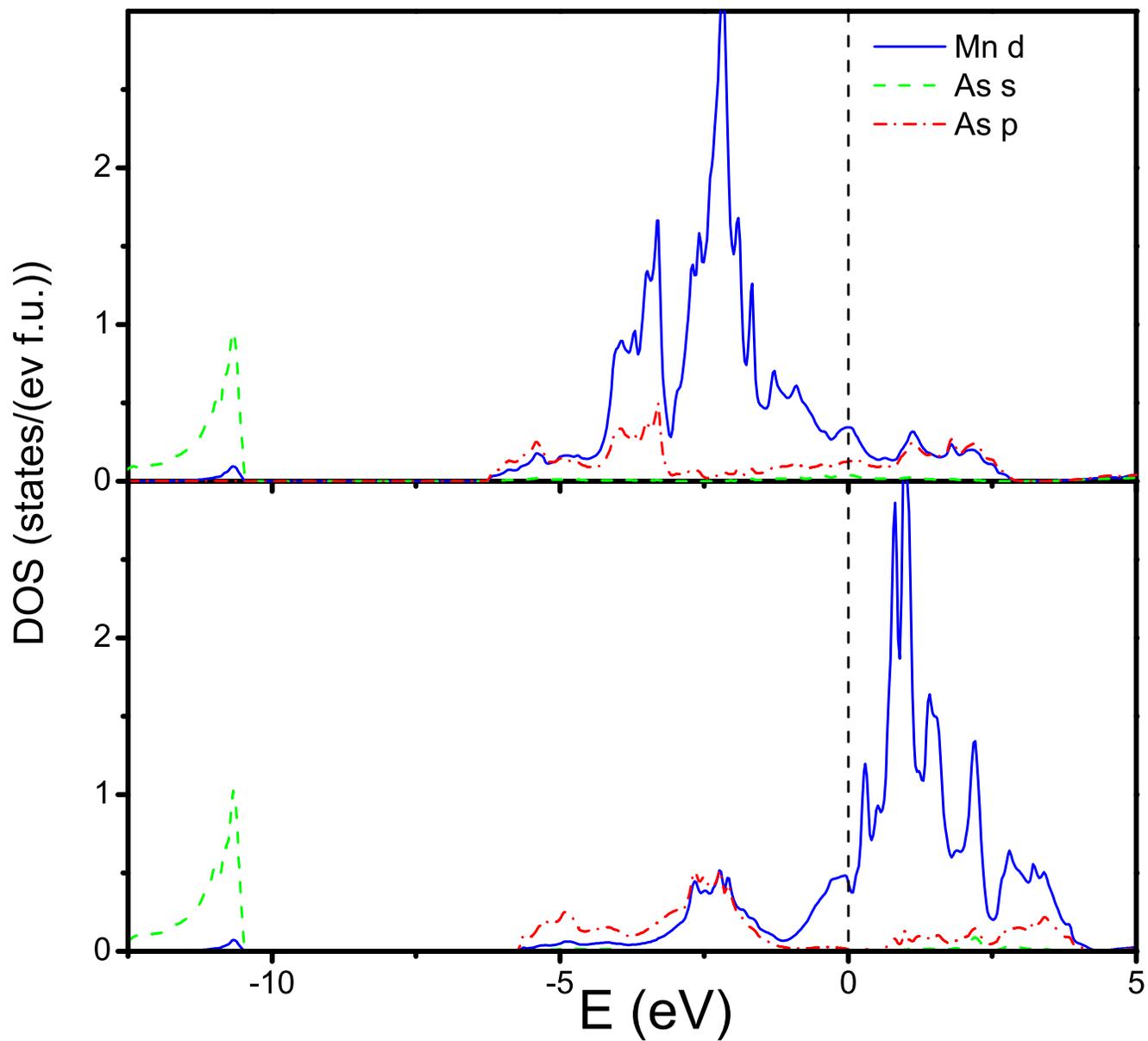}
\caption{\label{spd}(Color online) Density of states of MnAs in the 
hexagonal phase divided in the most important contributions 
for the majority (upper figure) and minority (lower figure) states.}
\end{figure}

The spin projected density of states (DOS) for the three manganese 
pnictides is shown in figure \ref{DOS}, with energy reference 
equal to the Fermi energy. The band structure has been obtained before 
for these compounds by several authors. % The non-magnetic DOS of MnAs 
% and MnSb in the NiAs and MnP type structures have been calculated by 
% Motizuki et al.~\cite{Motizuki_JPC_85}, using the old APW method with 
% muffin tin potentials, and the LDA. 
Although the DOS obtained by~\cite{MnAs:Rav99} is different, coincidentally 
most of the qualitative features apply also.
Mainly Mn d and pnictide p states hybridize decreasing the free value $5\mu_B$ .
%In our case the hybridization is higher, resulting in a lower magnetic moment.
In both our study and theirs there are large peaks at 
the up states, below the Fermi energy,
largely due to Mn d states, at approximately -2.5 eV.
The Mn d-states for the down spin are shifted to the conduction 
band. This can be seen in figure~\ref{spd}, where the important 
states of each atom are discriminated in their s, p and d 
character. The As s states are nearly isolated between 13 and 
10.5 eV below the Fermi energy. The total number 
of states at the Fermi level is
2.51 for MnAs, 2.22 for MnSb, and 1.95 for MnBi.
In comparison, with the anti structure the values are 
higher, respectively 3.46, 2.78, 2.05~\cite{MnAs:Rav99},
which suggests that the structure is not so stable, as expected .
The experimental value of $2.4\pm0.4$ states for MnSb 
estimated from specific heat measurements~\cite{Liang1977} is also in agreement.

The magnetic hyperfine field increases greatly from MnSb to MnBi,
due to the larger polarization from the $s$ electrons at the nuclear position,
 which largely increases due to the additional s-orbitals of 
higher principal atomic number, 
and the fact that Mn in MnBi
has the larger magnetic moment, so that it polarizes the Bi valence electrons.
The magnetic moment of the pnictogen site is very small, so that the
core contribution of the hyperfine field is also small. 
For MnSb, previous NMR measurements have determined 
a frequency of $260$ MHz at low temperatures, attributed 
to domain wall edge resonances~\cite{NarasimhaRao1985}, correspondig 
to a hyperfine field  $B_{hf}=3.93$ T, equal to our calculated 
value for the bulk.
The hyperfine field has been measured in MnBi, at the Bi atoms 
by nuclear orientation~\cite{Koyama1977}, B$_{hf}=94$ T, comparing 
reasonably with our value of 81.8 T.

%%%%%%%%%%%%%%%%%%%%%%%%%%%%%%%%%%%%%%%%%%%%%%%%%%%%%%%%%%%%%%%%%%%%%%%%%%
%%%%%%%%%%%%%%%%%%%%%%%%%%%%%%%%%%%%%%%%%%%%%%%%%%%%%%%%%%%%%%%%%%%%%%%%%%
\section{Conclusion}
%%%%%%%%%%%%%%%%%%%%%%%%%%%%%%%%%%%%%%%%%%%%%%%%%%%%%%%%%%%%%%%%%%%%%%%%%%
%%%%%%%%%%%%%%%%%%%%%%%%%%%%%%%%%%%%%%%%%%%%%%%%%%%%%%%%%%%%%%%%%%%%%%%%%%
We have measured the hyperfine parameters with the 
perturbed angular correlation method in MnAs.
The hysteresis at the hexagonal-orthorhombic 1st order 
phase transition is clearly seen from a microscopic point 
of view, complementing the X-ray and magnetization measurements.
The hyperfine magnetic field is the same at a given temperature, irrespective 
of cooling or heating the sample even in the phase coexistence 
region. 
This local probe study shows 
that the magnetization changes observed are mostly due 
to a change of phase fractions, which can be related to XRD 
studies. We provide a clear demonstration of the nature of the  first-order 
phase transition, by microscopic observation of phase separation at the 
hyperfine interactions range (sentitive to approximately less than $10$ \AA{}), much 
shorter than the range of 
diffraction techniques. 
We measured phase coexistence in a small interval 
of temperature ($2$ C), comparable with previous measurements.
In contrast, in other cases, PAC measurements were able to find very small coexistent 
regions of two competing phases, in a much broader temperature range 
than that given by x-ray diffraction~\cite{Lopes2006}.

Ab-initio calculations are used to complement the experiment. 
Realistic simulations of the diluted probe with supercell 
calculations show that the $^{77}$Se probe, if 
substitutional, is located at the As site. 
This information is taken from the 
comparison 
of calculated and measured hyperfine parameters, 
and it is verified by the 
calculated formation energies. Our results reproduce 
the hyperfine field at low temperature with good quantitative agreement. 

It may be 
interesting to try an experiment with a probe of high quadrupole 
moment, since in this case the EFG is very small and 
 has an overlying high amplitude magnetic hyperfine field,
which makes its accurate determination difficult. Improved 
results for the compounds MnSb and MnBi, of hyperfine 
parameters, magnetic moments and density of states 
were also presented and discussed. 

%We have measured the hyperfine parameters with the 
%perturbed angular correlation method in MnAs.
%The hysteresis at the hexagonal-orthorhombic 1st order 
%phase transition is clearly seen from a microscopic point 
%of view, complementing the macroscopic X-ray and magnetization measurements.
%The HMF is the same at a given temperature, irrespective 
%of cooling of heating the sample even in the phase coexistence 
%region. These results are also consistent with X-ray and 
%magnetization measurements. This local probe study shows 
%that the magnetization changes observed are mostly due 
%to a change of phase fractions, which can be related to XRD 
%studies. We measured phase coexistence in a small interval 
%of temperature (less than 2 C).
%Ab-initio calculations show that the $^{77}$Se probe, if 
%substitutional, is located at the As site. The theory reproduces 
%the hyperfine field at low temperature with good agreement, while 
%at room temperature the calculated field is too high. It may be 
%interesting to try an experiment with a probe of high quadrupole 
%moment, since in this case the EFG is very small and 
% has an overlying high amplitude magnetic hyperfine field,
%which makes its accurate determination difficult. Improved 
%results for the compounds MnAs, MnSb, and MnBi, of hyperfine 
%parameters, magnetic moments and density of states 
%were also presented and discussed. 

\begin{acknowledgments}
This work was supported by the Portuguese Foundation for Science 
and Technology FCT, with projects CERN-FP-109357-2009, CERN-FP-109272-2009, 
the German BMBF funding resources and by the ISOLDE collaboration with
approved project IS487. J. N. Gon\c{c}alves acknowledges
FCT PhD grant SFRH/BD/42194/2007. The authors
gratefully acknowledge S. Gama for supplying the MnAs samples, R. Soares 
for XRD measurements
 and H. Haas for useful discussions.
\end{acknowledgments}

%%%%%%%%%%%%%%%%%%%%%%%%%%%%%%%%%%%%%%%%%%%%%%%%%%%%%%%%%%%%%%%%%%%%%%%%%%%%%%%%
%%%%%%%%%%%%%%%%%%%%%%%%%%%%%%%%%%%%%%%%%%%%%%%%%%%%%%%%%%%%%%%%%%%%%%%%%%%%%%%%
%%%%%%%%%%%%%%%%%%%%%%%%%%%%%%%%%%%%%%%%%%%%%%%%%%%%%%%%%%%%%%%%%%%%%%%%%%%%%%%%
%%%%%%%%%%%%%%%%%%%%%%%%%%%BIBLIOGRAPHY%%%%%%%%%%%%%%%%%%%%%%%%%%%%%%%%%%%%%%%%%
%
%%%%%%%%%%%%%%%%%%%%%%%%%%%%%%%%%%%%%%%%%%%%%%%%%%%%%%%%%%%%%%%%%%%%%%%%%%%%%%%%%
%%%%%%%%%%%%%%%%%%%%%%%%%%%%%%%%%%%%%%%%%%%%%%%%%%%%%%%%%%%%%%%%%%%%%%%%%%%%%%%%%
%%%%%%%%%%%%%%%%%%%%%%%%%%%%%%%%%%%%%%%%%%%%%%%%%%%%%%%%%%%%%%%%%%%%%%%%%%%%%%%%%
%%%%%%%%%%%%%%%%%%%%%%%%%%%%%%%%%%%%%%%%%%%%%%%%%%%%%%%%%%%%%%%%%%%%%%%%%%%%%%%%%
%%%%%%%%%%%%%%%%%%%%%%%%%%%%%%%%%%%%%%%%%%%%%%%%%%%%%%%%%%%%%%%%%%%%%%%%%%%%%%%%%
%%%%%%%%%%%%%%%%%%%%%%%%%%%%%%%%%%%%%%%%%%%%%%%%%%%%%%%%%%%%%%%%%%%%%%%%%%%%%%%%%
\begin{table}\centering
\caption{\label{fitparup}Fit parameters, from measurements when 
heating and cooling the 1st sample, in chronological order. 
Fraction (\%) of the measured interactions, quadrupolar 
frequency ($\omega_{0}$) (Mrad.s$^{-1}$), 
Larmor frequency $\omega_L$ (Mrad.s$^{-1}$)
and width of Lorentzian function used in the fitting 
procedure ($\sigma$) (Mrad.s$^{-1}$). 
The magnetic frequencies have an estimated upper limit 
for the error of 15 Mrad.s$^{-1}$. 
The  fraction $f_3$ and 
the frequency of the orthorhombic phase were kept 
fixed at an average value.  H = hexagonal 
ferromagnetic, O = orthorhombic, H$^*$ = hexagonal paramagnetic.}
\begin{tabular}{|c|c|c|c|c|c|c|c|c|c|c|}
\hline\hline
$T(^{\circ}$C$)$     &
$f_1$ & $\omega_{L1}$& $\sigma_1$       &
$f_2$& $\omega_{02}$& $\sigma_2$  &
$f_3$& $\omega_{03}$&$\sigma_3$& Phase\\
\hline
17& 78 & 637 & 21&-&-&-&22& 945& 69  &             \\
35& 78 & 547 & 28&-&-&-& 22 & 1466  & 69&H\\
41.5& 78 & 506 & 33&-&-&-& 22 & 1484  & 55&\\
\hline
50&-&-&-& 78 & 29& 290& 22& 1108& 10 &\\
45&-&-&-& 78& 29&  330& 22& 874& 96 &O\\
41&-&-&-& 78& 29& 290& 22& 963& 46 &\\
\hline
141 &-&-&-& 78 & 29&  435& 22& 1324& 17&H$^*$\\
\hline
21.2& 78& 618 & 33&-&-&-& 22& 176& 40&H \\
\hline
35& 60& 546 & 65&18&28&31& 22& 176& 43& H+O\\
\hline\hline
\end{tabular}
\end{table}

\begin{table}\centering
\caption{\label{fitpar2}Fit parameters, with temperatures 
in chronological order, for the 2nd sample. Fraction (\%) of the measured 
interactions, quadrupolar frequency $\omega_{0}$ (Mrad.s$^{-1}$), 
Larmor frequency $\omega_L$ (Mrad.s$^{-1}$), and 
width of Lorentzian function used in the fitting 
procedure $\sigma$ (Mrad.s$^{-1}$). 
The magnetic frequencies have an estimated upper 
limit for the error of 15 Mrad.s$^{-1}$. 
$f_3$ and corresponding frequency were fixed at an 
average value, as was $\omega_{02}$. H = hexagonal 
ferromagnetic, O = orthorhombic.}
\begin{tabular}{|c|c|c|c|c|c|c|c|c|c|c|}
\hline\hline
$T(^{\circ}$C$)$     &
$f_1$& $\omega_{L1}$&  $\sigma_1$       &
$f_2$& $\omega_{02}$&  $\sigma_2$ &
$f_3$& $\omega_{03}$&   $\sigma_3$&\\
\hline
21.1& 70 & 619& 3&-&-&-& 30& 176&  23  & \\
40.8& 70 & 510& 9&-&-&-& 30& 176&   22& H\\
41.3& 70 & 499& 12&-&-&-& 30& 176&  18 & \\
\hline
42.3& 40 & 495& 10&30&28&20& 30& 176&  10 & H+O \\
\hline
43.5&-&-&-& 70 & 28 &  24& 30& 176&  9 & \\
124.5&-&-&-&70& 28 &  26& 30& 176&  27 & \\
41.4&-&-&-& 70 & 28 &  36& 30& 176&  17& \\
39.5&-&-&-& 70 & 28 &  46& 30& 176&  24& O\\
37.5&-&-&-& 70 & 28&  47& 30& 176&  22 & \\
34.5&-&-&-& 70 & 28 &  29& 30& 176&  14& \\
33.3&-&-&-& 70 & 28 & 197& 30& 176&   73  & \\
32.5&-&-&-& 70 & 28 &  20& 30& 176& 11 & \\
\hline
29.9& 70 &  571& 9&-&-&-& 30& 176&  48 & \\
13.6& 70 &  648& 4&-&-&-& 30& 176&  98 & H\\
-196& 70 &  1050& 0&-&-&-& 30& 57&  29 & \\
\hline\hline
\end{tabular}
\end{table}

\begin{table}[ht]\centering
\caption{\label{EFGAs}Calculated electric field gradient of MnAs, at 
the hexagonal phase, with room temperature lattice 
constants: a=3.722 \AA{}, c=5.702 \AA{}.}
\begin{tabular}{lrrr}
\hline\hline
Atom                 & V$_{zz}$($10^{21}$V/m$^2$)    & $\eta$    & V$_{zz}$ dir. \\
\hline
Mn &                 -3.63         &0         &(0,0,1)\\
As &                 1.53         &0        &(0,0,1)\\
\hline\hline
\end{tabular}
\end{table}

\begin{table}[ht]\centering
\caption{\label{BHF1}Calculated hyperfine Fields of MnAs (T): room 
temperature lattice constants: a=3.722 \AA{}, c=5.702 \AA{}, hexagonal phase.}
\begin{tabular}{lrrrrr}
\hline\hline
Atom                & $B_C$     & core    & valence & $B_{orb}$ & $B_{dip}$\\
\hline
Mn & -6.5 & -39.3 & 32.8 &0.5&-3.0\\
As & 25.0 & 0.4 & 24.5 &-0.1&-0.2\\
\hline\hline
\end{tabular}
\end{table}

\begin{table}[ht]\centering
\caption{\label{BHF2}Calculated contact hyperfine 
field of MnAs (T): low temperature lattice 
constants: a=3.732 \AA{}, c=5.678 \AA{}, hexagonal phase.}
\begin{tabular}{lrrr}
\hline\hline
Atom               & $B_C$     & core   & valence \\
\hline
Mn & -6.1 & -38.7 & 32.5 \\
As & 25.5 & 0.4 & 35.1\\
\hline\hline
\end{tabular}
\end{table}

\begin{table}[ht]\centering
\begin{tabular}{rrrrrr}
\hline\hline
Se at As site\\
\hline
V$_{zz}$($10^{21}$V/m$^2$)&$\eta$&V$_{zz}$ dir. \\
-0.27         & 0         & (0,0,1)\\
$B_C$      & core   & valence&$B_{orb}$&$B_{dip}$&||$B_{hf}$ total (T) \\
56.6          & 0.4& 56.2&-2.1&-0.2&54.3 \\
Se at Mn site\\
\hline
V$_{zz}$($10^{21}$V/m$^2$)&$\eta$& V$_{zz}$ dir. \\
17.80         & 0         & (0,0,1)\\
B$_C$      & core   & valence &$B_{orb}$&$B_{dip}$&|| $B_{hf}$ total (T)\\
-23.1        & -22.9 & -0.2 &3.4&-1.8&-21.5\\
\hline\hline
\end{tabular}
\caption{\label{EFGSe}Hyperfine parameters with Se 
probe substitutional at the As or Mn sites in MnAs.}
\end{table}

\begin{table}[ht]\centering
\caption{\label{EFGpnictides}Electric field gradient of 
MnAs, MnSb and MnBi, p-p and d-d contributions in the atomic spheres.}
\begin{tabular}{lrrrr}
\hline\hline
Compound&Atom&V$_{zz}$($10^{21}$V/m$^2$)&V$_{zz}^{p-p}$&V$_{zz}^{d-d}$\\
MnAs&Mn &                 -3.63      & -1.45 & -1.70\\
&As &                 1.53         & 1.27& 0.06\\
MnSb&Mn &                 -3.92&-1.40&-1.97\\
&Sb &                 4.67&3.97&0.10\\
MnBi&Mn &                 -4.43        &-1.91&-2.49\\
&Bi &                 9.46        &9.44&0.21\\
\hline\hline
\end{tabular}
\end{table}

\begin{table}[ht]\centering
\caption{\label{MM} Total magnetic moment for MnX (X=P, As, Sb, Bi) in the
 cell and magnetic moments inside the Mn and X LAPW spheres, 
in units of $\mu_B$/formula unit. Previous experiments and 
calculations are compared with our results.}
\begin{tabular}{cccc}
\hline\hline$10^{21}$V/m$^2$
Compound&Mn$_s$ &X$_s$ &MnX$_t$ \\
\hline
MnAs(present work) &   3.29&-0.14&3.17 \\
MnAs(exp.) &&-0.23~\cite{Yamaguchi35}&3.40~\cite{Yamaguchi35} \\
MnAs(theory) &&&3.10~\cite{Katoh36}\\
 &3.14&-0.08&3.06~\cite{Oppeneer37}\\
 &3.18&-0.13&3.14~\cite{jamal}\\
\hline
MnSb(present work) &   3.44&-0.14&3.34\\
MnSb(exp.) &&-0.30~\cite{Yamaguchi35}&3.55~\cite{Bouwma1971}; 3.50~\cite{Chen39}\\
MnSb(theory) &   3.34&-0.07&3.27~\cite{Katoh36}\\
 &   3.35&-0.032&3.32~\cite{Vast40}\\
 &   3.30&-0.06&3.24~\cite{Coehoorn41}\\
 &   3.5&-0.17&3.31~\cite{Podoloucky42}\\
\hline
MnBi(present work) &   3.49&-0.11&3.42\\
MnBi(exp.) &&&3.82~\cite{Zhiqiang43}; 3.84~\cite{Chen44}; 3.9~\cite{Heikes45}\\
MnBi(theory) &3.71&-0.10&3.61~\cite{Oppeneer37}\\
 &3.50&-0.02&3.56~\cite{Kohler46} \\
\hline\hline
\end{tabular}
\end{table}

\begin{table}\centering
\caption{\label{BHFpnictides}Fermi contact hyperfine 
Fields of MnAs, MnSb, and MnBi, core and valence 
contributions, orbital and dipolar hyperfine fields ($T$).}
\begin{tabular}{llrrrr}
\hline\hline
Atom & $B_C$ & core & valence & $B_{orb}$ & $B_{dip}$ \\
\hline
Mn & -6.5 & -39.3 & 32.8 & 0.5 & -3.0 \\
As & 25.0 & 0.4 & 24.6 &-0.1 & -0.2\\
\hline
Mn & -8.5 & -41.5 & 33.0 & 1.2 &-2.8\\
Sb & 30.6 & -2.0 & 30.8 & 0.1 & -0.5\\
\hline
Mn & -5.8 & -42.1 & 36.3 & 3.6&-2.8 \\
Bi & 82.6 & -1.2 & 83.8 & 0.2&-1.0 \\
\hline\hline
\end{tabular}
\end{table}

%%%%%%%%%%%%%%%%%%%%%%%%%%%%%%%%%%%%%%%%%%%%%%%%%%%%%%%%%%%%%%%%%%%%%%%%%%%%%%%%%
%%%%%%%%%%%%%%%%%%%%%%%%%%%%%%%%%%%%%%%%%%%%%%%%%%%%%%%%%%%%%%%%%%%%%%%%%%%%%%%%%
%%%%%%%%%%%%%%%%%%%%%%%%%%%%%%%%%%%%%%%%%%%%%%%%%%%%%%%%%%%%%%%%%%%%%%%%%%%%%%%%%
%%%%%%%%%%%%%%%%%%%%%%%%%%%%%%%%%%%%%%%%%%%%%%%%%%%%%%%%%%%%%%%%%%%%%%%%%%%%%%%%%
%%%%%%%%%%%%%%%%%%%%%%%%%%%%%%%%%%%%%%%%%%%%%%%%%%%%%%%%%%%%%%%%%%%%%%%%%%%%%%%%%
%%%%%%%%%%%%%%%%%%%%%%%%%%%%%%%%%%%%%%%%%%%%%%%%%%%%%%%%%%%%%%%%%%%%%%%%%%%%%%%%%

\end{document}